\documentclass[aps,prb,twocolumn,showpacs,superscriptaddress]{revtex4-1}
\usepackage[pdftex]{graphicx}
\usepackage{amssymb,amsmath}
\usepackage{dcolumn}
\usepackage{bm}
\usepackage{color,soul}

\begin {document}

\title{Observation of a temperature dependent asymmetry in the domain structure of a Pd doped FeRh epilayer}
\author{C.~J.~Kinane}
\email[Electronic mail: ]{christy.kinane@stfc.ac.uk}
\affiliation{ISIS, Rutherford Appleton Laboratory, Harwell Science
and Innovation Campus, Science and Technology Facilities Council,
Oxon. OX11 0QX, United Kingdom}
\author{M.~Loving}
\affiliation{Department of Chemical Engineering, Northeastern
University, Boston, MA 02115 USA}
\author{M.~A.~de~Vries}
\altaffiliation{Current Address: School of Chemistry, The University
of Edinburgh, Edinburgh, EH9 3JJ. United Kingdom}
\affiliation{School of Physics and Astronomy, University of Leeds,
Leeds LS2 9JT, United Kingdom}

\author{R.~Fan}
\altaffiliation{Current Address: Diamond Light Source Ltd, Harwell Science and
Innovation Campus, Didcot, Oxon, OX11 0DE, United Kingdom}
\affiliation{ISIS, Rutherford Appleton Laboratory, Harwell Science
and Innovation Campus, Science and Technology Facilities Council,
Oxon. OX11 0QX, United Kingdom}

\author{T.~R.~Charlton}
\affiliation{ISIS, Rutherford Appleton Laboratory, Harwell Science
and Innovation Campus, Science and Technology Facilities Council,
Oxon. OX11 0QX, United Kingdom}
\author{J.~S.~Claydon}
\affiliation{School of Physics and Astronomy, University of Leeds,
Leeds LS2 9JT, United Kingdom}
\author{D.~A.~Arena}
\affiliation{National Synchrotron Light Source, Brookhaven National
Laboratory, Upton, New York 11973-5000, USA}
\author{F.~Maccherozzi}
\author{S.~S.~Dhesi}
\affiliation{Diamond Light Source Ltd, Harwell Science and
Innovation Campus, Didcot, Oxfordshire OX11 0DE, United Kingdom}
\author{D.~Heiman}
\affiliation{Department of Physics, Northeastern
University, Boston, MA 02115 USA}
\author{C.~H.~Marrows}
\affiliation{School of Physics and Astronomy, University of Leeds,
Leeds LS2 9JT, United Kingdom}
\author{L.~H.~Lewis}
\affiliation{Department of Chemical Engineering, Northeastern
University, Boston, MA 02115 USA}
\author{Sean~Langridge}
\affiliation{ISIS, Rutherford Appleton Laboratory, Harwell Science
and Innovation Campus, Science and Technology Facilities Council,
Oxon. OX11 0QX, United Kingdom}
\date{\today}

\begin{abstract}
  Using X-ray photoelectron emission microscopy we have observed the coexistence of ferromagnetic and antiferromagnetic phases in a (3 at.\%)Pd-doped FeRh epilayer. By quantitatively analyzing the resultant images we observe that as the epilayer transforms there is a change in magnetic domain symmetry from predominantly twofold at lower temperatures through to an equally weighted combination of both four and twofold symmetries at higher temperature. It is postulated that the lowered symmetry Ising-like nematic phase resides at the near-surface of the epilayer. This behavior is different to that of undoped FeRh suggesting that the variation in symmetry is driven by the competing structural and electronic interactions in the nanoscale FeRh film coupled with the effect of the chemical doping disorder.
\end{abstract}


\maketitle

\section{Introduction}

The binary alloy FeRh exhibits a fascinating first-order transition from an antiferromagnetic (AFM) to a ferromagnetic (FM) state around
400 K\cite{kouvel1962,Lommel1966}. This transition is accompanied by a significant magnetoresistance\cite{algarabel1995,thiele2003,Kushwaha2009,Suzuki2011}, a large lattice expansion\cite{zsoldos1967,Lbarra1994} and entropy release\cite{thiele2003}. The ability to produce epitaxial thin films of FeRh has revealed additional complexity with a surface related FM state present in the nominally AFM phase\cite{Fan2010,Suzuki2009}. These results have been confirmed by near-surface sensitive real-space imaging performed using soft X-ray photoelectron emission microscopy (XPEEM) which show significantly different behavior for capped and un-capped samples\cite{Baldasseroni2012,Baldasseroni2014}. Temperature-dependent, hard X-ray photoemission spectroscopy has demonstrated changes in the core Fe $2p$ levels and in the valence band structure in remarkable agreement with results obtained from density functional theory\cite{Gray2012,Sandratskii2011} and show that the metamagnetic transition is likely to be driven by an electronic transition. $^{57}$Fe conversion electron M\"{o}ssbauer spectroscopy has also revealed a strain-driven reorientation of the spins at the AFM-FM phase transition in FeRh thin films\cite{PhysRevLett.109.117201}. Hall-effect measurements across the metamagnetic transition are also consistent with an electronic transition leading to a large increase in the carrier density in the FM phase. This effect has recently been utilised in a room temperature controllable resistor\cite{Marti2014} making use of the anisotropic magnetoresistance
(AMR) in the antiferromagnetic state of the FeRh. Furthermore, FeRh has attracted significant interest due to its ultrafast dynamics in which there is much debate over the out-of-equilibrium state\cite{Quirin2012,thiele_spin_2004,Radu2010}.

The magnetic behavior of FeRh is strongly affected by doping with other transition metals\cite{Miyajima_JJAP_1993,Yuasa_JPSJ_1995,Yuasa_NIMB_1993}. The AFM to FM transition temperature can be increased by doping with Ir and Pt and decreased by doping with Pd and Ni. This allows the AFM to FM transition temperature to be tuned down to room temperature\cite{Baranov1995139,kouvel:1257,walter:938}. Doping FeRh with Pd (FeRh$_{1-x}$Pd$_{x}$) has been shown, in bulk samples, to preserve the B2 CsCl structure. However, the $c/a$ ratio increases linearly with increased Pd doping\cite{Yuasa_JPSJ_1995,Miyajima_JJAP_1993}. The Pd is known to substitute onto the Rh sites\cite{Yuasa_JPSJ_1995} and continues up to the level $x\approx$0.3 where an AFM-paramagnetic transition replaces the AFM-FM transition. It is also possible to tune the FeRh transition via pressure\cite{JPSJ.63.855} and magnetic field\cite{Maat2005}. In the case of pressure tuning, the transition temperature increases by $\approx 5$~K/kbar\cite{Vinokurova1981}, while for field tuning it decreases by $\approx 8$~K/T. Recently the injection of spin-polarized current\cite{Naito2011} has been shown to promote the transition in FeRh from AFM to FM.

As the FeRh transition is thermodynamically first-order there is expected to be a phase coexistence as the system transforms. In thin films the structural phase coexistence is clearly evident from the observation of a well defined change in the out-of-plane lattice parameter and a small change in the in-plane lattice parameters associated with the differing unit cell sizes of the AFM and FM phases\cite{kim2009,Fan2010}. Furthermore, in the transition region there is expected to be a magnetic phase coexistence of AFM and FM regions. This phase coexistence has been observed using X-ray diffraction\cite{Vries_APL_2014} along with a more distinct separation of the phases on cooling through the transition than on warming. This is found to be consistent with a melting/freezing first order phase transition.

The coexistence of AFM and FM regions as well as the presence of structural and electronic competition in FeRh is similar to the mesoscopic phase separation observed in strongly correlated oxides at low temperature\cite{Mathur2003,Dagotto2005}. In such systems, competing degrees of freedom result in charge, orbital and magnetic phase separation in otherwise chemically homogeneous samples, resulting in a rich phase diagram. Complex oxide behavior can similarly be tuned via doping that acts as a source of quenched disorder and produces randomly arranged (sub-)micron sized domains of charge order and ferromagnetism\cite{Moreo2000,Uehara1999}. As the phase transition in FeRh is accompanied by a change in the sign and magnitude of the magnetic exchange coupling, a large structural modification, and an electronic transition\cite{Kushwaha2009,Gray2012,Vries2013} it is interesting to consider whether a similar degree of complexity is observed in a binary alloy at room temperature.

In this report, we image the magnetic phase coexistence in Pd doped FeRh thin films and observe an unusual Ising-like nematic ordering upon warming into the FM state from the AFM state. A stable and controllable domain structure in the hysteretic regime with electronic and magnetic phase coexistence could be harnessed for novel functionalities\cite{Cherifi2014}, for example in memory cells. For such applications it would be highly desirable for the hysteretic regime to be centered around room temperature.

\section{Sample preparation and characterization}

\begin{figure}
  \includegraphics[width=60mm]{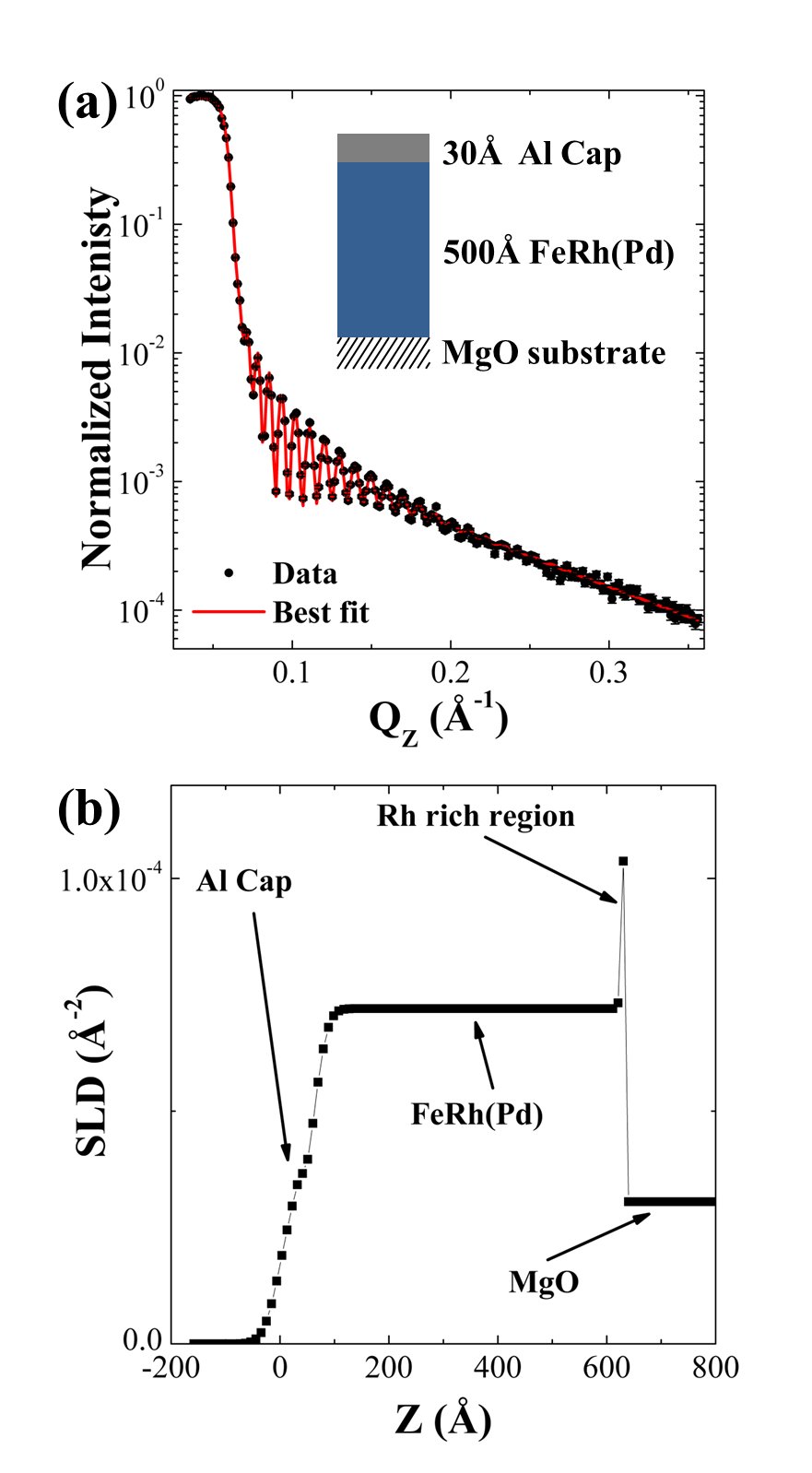}
  \caption{Structural characterization of the FeRh(Pd) epilayer:(a) XRR from the epilayer. The best fit to the data is shown as the solid curve. The inset shows a schematic of the nominal sample stack. (b) The X-ray scattering length density profile extracted from the best fit of the data in panel (a). }\label{Fig1}
\end{figure}

The sample studied is an epilayer grown by DC magnetron sputtering on MgO single crystal substrates according to the methods described in Ref. \onlinecite{graet_jove}. In this work we have doped the FeRh with 3 at.\% Pd which conveniently pushes the transition temperature down close to room temperature\cite{Kushwaha2009}. The film was co-deposited from separate angled Fe and Rh sources. Pd doping was provided by a small strip of Pd placed on the surface of the Rh target. The pressure during growth was $5 \times 10^{-8}$~Torr and the substrate temperature was 600$^\circ$C. Ar gas with 4\% H$_{2}$ at 3~mTorr was used as the sputter gas. The film was post-growth annealed at 700$^\circ$C for 60 minutes, and then cooled to 100$^\circ$C before being capped \textit{in-situ} with 30~\AA\ of Al. A schematic of the structure is shown in the inset of Figure~\ref{Fig1}(a). The FeRh composition was determined by energy dispersive X-ray (EDX) spectroscopy, on a $\approx100$~nm lamella prepared using focussed ion beam techniques, as Fe(48 at.\%)/Rh (49 at.\%)/Pd(3 at.\%) with a 3\% error. Low-angle X-ray reflectivity (XRR) data shown in Figure~\ref{Fig1}(a) were used to determine the average structure of the epilayer. At low angles the technique is not sensitive to the crystallinity and the electron density depth profile over the whole sample is analyzed by an optical matrix method\cite{Parratt1954}. The resultant depth profile is shown in Figure~\ref{Fig1}(b). The substrate interface is sharp with a root mean squared roughness (rms) of approx 0.7~$\pm$~0.5~\AA\ with the surface/cap region significantly more diffuse. Unsurprisingly the Al cap has oxidized forming a less dense but thicker oxide layer. The FeRh(Pd) layer was found to be 563~$\pm$~5~\AA\ thick. An additional layer with a slightly higher electron density compared to bulk Rh with thickness of $\sim8$~\AA\ was introduced between the MgO substrate and FeRh(Pd) layer in order to fully describe the reflectivity data\cite{Loving2012a,Baldasseroni2014}. The fitting parameters are displayed in Table.~\ref{table 1}.

\begin{table}
  \begin{ruledtabular}
  \begin{tabular}{ c|ccc }
    Material & d~(\AA) & $\rho$ SLD~(\AA$^{-2}$) & $\sigma$ (\AA) \\
  \hline
  Al$_{2}$O$_{3}$ (cap) & 43 $\pm$ 3 & (3.4 $\pm\ 0.6)\times 10^{-5}$ & 23 $\pm$ 11 \\
  Al(cap) & 17 $\pm$ 4 & (2.2 $\pm\ 0.2)\times 10^{-5}$ & 7 $\pm$ 3\\
  FeRh(Pd 3\%) & 563 $\pm$ 5  & (7.2 $\pm\ 0.2)\times 10^{-5}$ & 20 $\pm$ 9 \\
  Rh  & 8.4 $\pm$ 1.5 & (1.0 $\pm\ 0.5)\times 10^{-4}$ & 1.4 $\pm$ 0.6 \\
  MgO(sub) & $\infty$  & (3.1 $\pm\ 0.1)\times 10^{-5}$ & 0.7 $\pm$ 0.5 \\
  \end{tabular}
  \end{ruledtabular}
  \caption{Table of fitted parameters obtained using the PARRAT32 software\cite{Braun1997} from the Cu K$_{\alpha}$ XRR data displayed in Figure \ref{Fig1}. $d$ is the film thickness, $\rho$ is the layer's scattering length density (SLD) and $\sigma$ the interfacial rms roughness.} \label{table 1}
\end{table}

\begin{figure}
  \includegraphics[width=80mm]{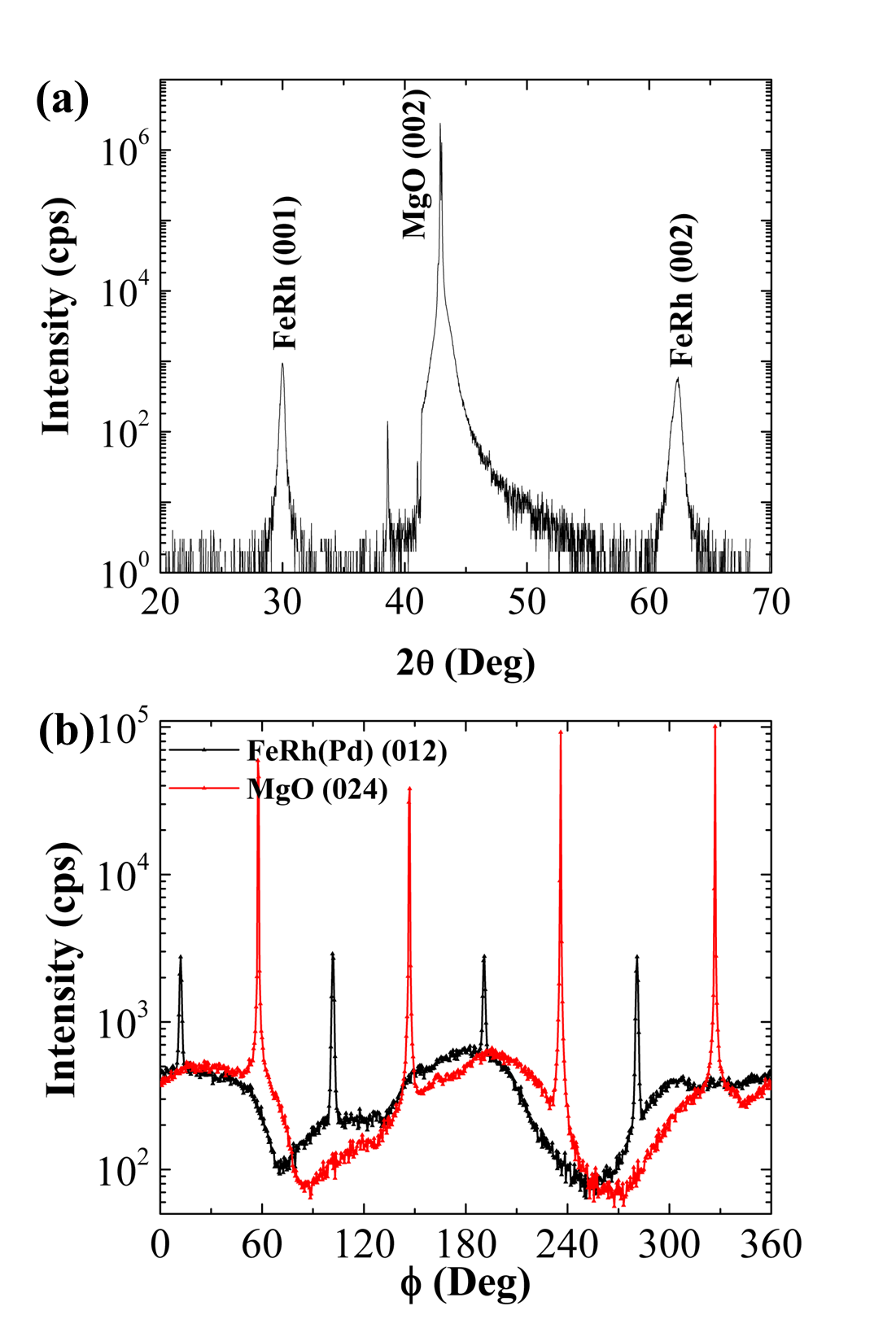}
  \caption{Structural characterization of the FeRh(Pd) epilayer and MgO substrate: (a) High angle Cu $K_{\alpha}$ XRD of the epilayer showing the highly ordered single crystalline nature of the Pd-doped FeRh film.(b) $\phi$-scans through the MgO (024) and FeRh (012) peaks, showing the expected fourfold symmetry.}\label{Fig1a}
\end{figure}

X-ray diffraction (XRD) data are shown in Figure~\ref{Fig1a}. Clear (001) and (002) diffraction  peaks of the highly-chemically-ordered FeRh phase with the B2 CsCl structure ($\alpha$$^\prime$ phase) are observed. The out-of-plane lattice constant at room temperature was calculated from the FeRh $(00L)$ peak positions and has a value of 2.998~\AA. This value matches the undoped sample $c=2.995$~\AA\ and compares favorably to the bulk value of 2.989~\AA\ reported by Lommel\cite{Lommel1966}. The MgO substrate is known to apply an in-plane compressive strain producing the observed out-of-plane lattice expansion\cite{Fan2010}. The long range order parameter \textbf{S} was determined using the procedure described by Warren\cite{Warren_xraydiff_1990}. It was found to be \textbf{S}$\approx$0.86 and in good agreement with previous work\cite{Vries2013}, considering the inclusion of Pd doping. The FeRh(Pd3\%) layer was confirmed by XRD to have the expected fourfold in-plane symmetry and the epitaxial relationship to the MgO substrate of FeRh(Pd)[001]$\|$MgO[001] and FeRh(Pd)[100]$\|$MgO[110]. The registry of the film and substrate is shown in Figure~\ref{Fig1a} (b).

\begin{figure*}
\includegraphics[width=170mm]{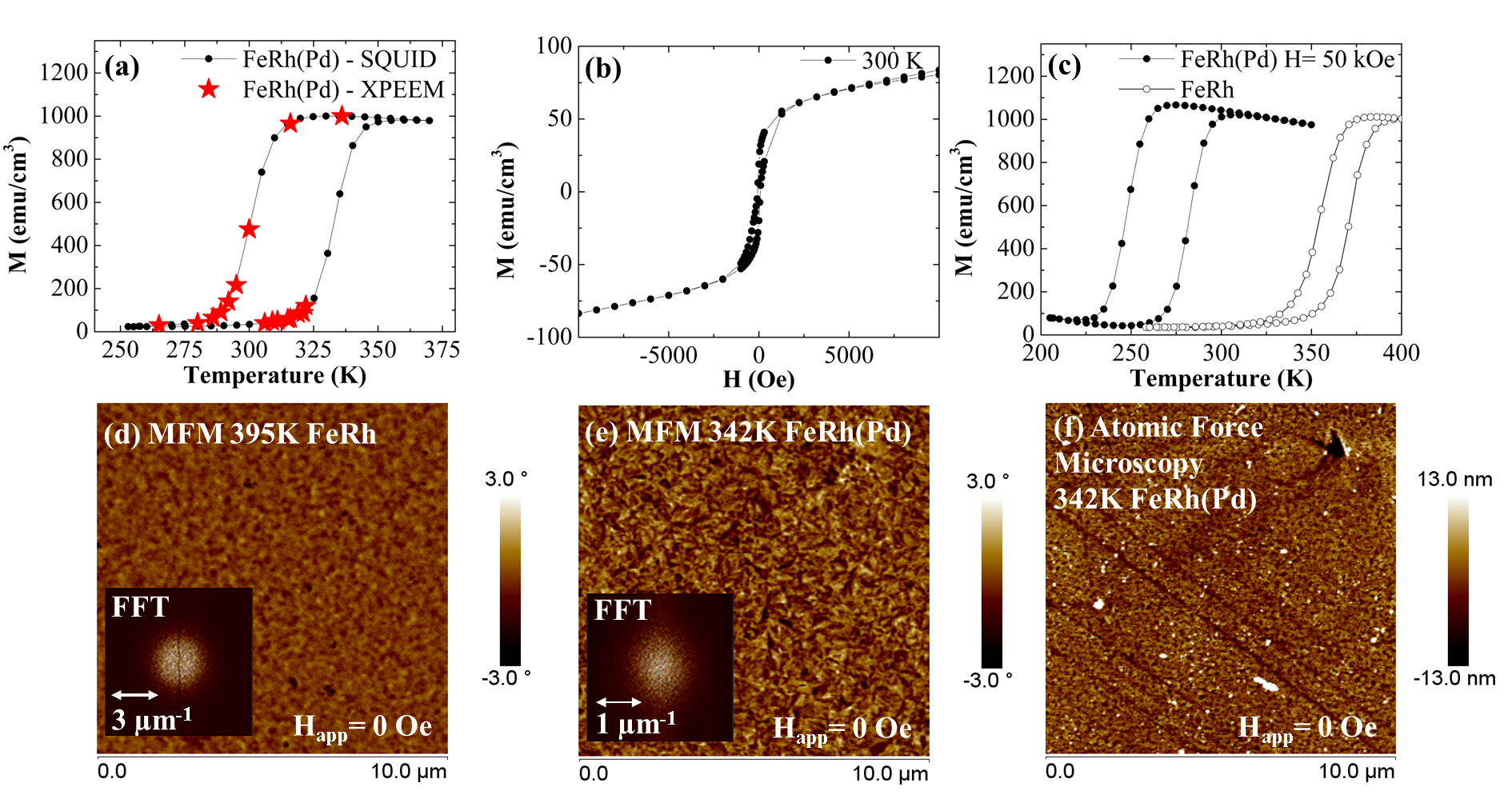}\\
\caption{Magnetic characterization of the FeRh(Pd) epilayer: (a) Magnetization versus temperature ($M(T)$) measured in a 100~Oe applied field for FeRh(Pd 3\%). The red stars correspond to the XPEEM measurement temperatures only. (b) Applied field dependence of the magnetization at 300~K showing a canted hysteresis loop. Panel (c) displays $M(T)$ for both the FeRh(Pd) and FeRh films in a field of 50~kOe to reduce the undoped FeRh transition into the measurable range of the SQUID. Note the similar magnitude of the magnetization and narrowing of the FeRh hysteresis. Panels (d) and (e) show MFM images of the FeRh (395~K) and FeRh(Pd)(342~K) films respectively at a temperature where the films are entering the phase coexistence regime upon warming. The inset in d) and e) show the FFTs of the images used to estimate the magnetic domain sizes. The images were measured in zero-field conditions. (f) Atomic force microscopy  topographic image of the surface of the FeRh(Pd) sample at 342~K. The diagonal grooves are scratches on the film surface.} \label{Fig2}
\end{figure*}

The sample's magnetic properties were studied using SQUID magnetometry and temperature-dependent magnetic force microscopy (MFM)(Bruker MultiMode 8 SPM, employing MESP probes). Figure~\ref{Fig2}(a) demonstrates that the MgO/FeRh(Pd)/Al film has a bulk-like transition from AFM to FM behavior upon heating to 300 K with a temperature hysteresis of about 30 K. Figure~\ref{Fig2}(b) shows the sample magnetization $M$ \textit{vs.} applied field $H$ measured at 300K on warming. The hysteresis loop qualitatively shows coexistence of both ferromagnetism and antiferromagnetism, having a coercive field of $\thickapprox$ 80~Oe and a canted hysteresis loop that is not fully saturated at 10~kOe, consistent with the undoped material. Figure~\ref{Fig2}(c) compares the thermal hysteresis loops of both FeRh(Pd) and pure FeRh measured in an applied field of 50~kOe, so as to shift the transition temperature into the measurement regime of our SQUID magnetometer. Both sample compositions have equivalent saturation moments; the doped system has a wider thermal hysteresis than does the pure system due to the increased disorder derived from the Pd doping.

Panels (d) and (e) of figure~\ref{Fig2} show 10~$\mu$m~$\times~10~\mu$m MFM images of FeRh at 395~K and FeRh(Pd)at 342~K respectively in zero applied magnetic field upon warming. At these temperatures both films are in the process of transforming from the AFM to the FM phase. MFM is inherently sensitive to out-of-plane magnetization, hence in the case of in-plane magnetization it is largely sensitive to domain walls and out-of-plane stray fields. A clear difference is observed between the undoped and doped systems: the undoped FeRh film exhibits isotropic magnetic domains\cite{Baldasseroni2012} and the Pd-doped FeRh film exhibits smaller, slightly elongated domains. Panel (f) of figure~\ref{Fig2} shows the atomic force microscopy topography of the FeRh(Pd) sample imaged at 342~K, hinting at pitting and scratches on the surface at a sub-micron length scale that is smaller than the magnetic domains. This feature indicates that the surface magnetism is largely decoupled from the surface topography in both films. The inset shows the fast Fourier transform (FFT) of the MFM image at 342~K in Figure~\ref{Fig2}(e) yielding a magnetic domain size of approximately 1.4~$\mu$m which also appears to be slightly asymmetric, while the FFT of the undoped FeRh sample, at an equivalent warming temperature (395~K), gives a larger domain size of 3.8~$\mu$m and is isotropic.

\section{XPEEM imaging}

The in-plane magnetic domain structure of the Pd-doped FeRh film was obtained using X-Ray Photoelectron Emission Spectroscopy situated on the Nanoscience Beamline I06 at the Diamond Light Source. In XPEEM measurements the magnetic contrast is obtained through the X-ray magnetic circular dichroism (XMCD) signal. Magnetism and element-specific images were acquired at the Fe L$_{III}$ absorption edge by exciting spin-polarized $2p$ core electrons into exchange-split unoccupied states above the Fermi level, and imaging the secondary electrons (total electron yield detection) in full-field mode. The beamline optics allow spot sizes of 10~$\mu$m to be generated and the final spatial resolution of the microscope is of order 100~nm. The magnitude of the dichroism observed in the XPEEM\cite{Schneider2002} is proportional to the cosine of the angle between the sample magnetization $\mathbf{M}$ and the direction of the photon helicity $\bm{\varepsilon}$. Hence XPEEM is sensitive to the magnitude of the vector component of the magnetization (anti)parallel to the direction of the photon propagation and is insensitive to the orthogonal vector components. The XMCD contrast can be extracted as a normalized difference of the observed photon helicity dependent intensities ($I^{\pm}$) defined as the spin asymmetry SA=(I$^{+}$-I$^{-}$)/(I$^{+}$+I$^{-}$). As a result, the strongest contrast is observed when the magnetization is aligned (anti)parallel to the photon propagation vector.

\begin{figure}
\includegraphics[width=70mm]{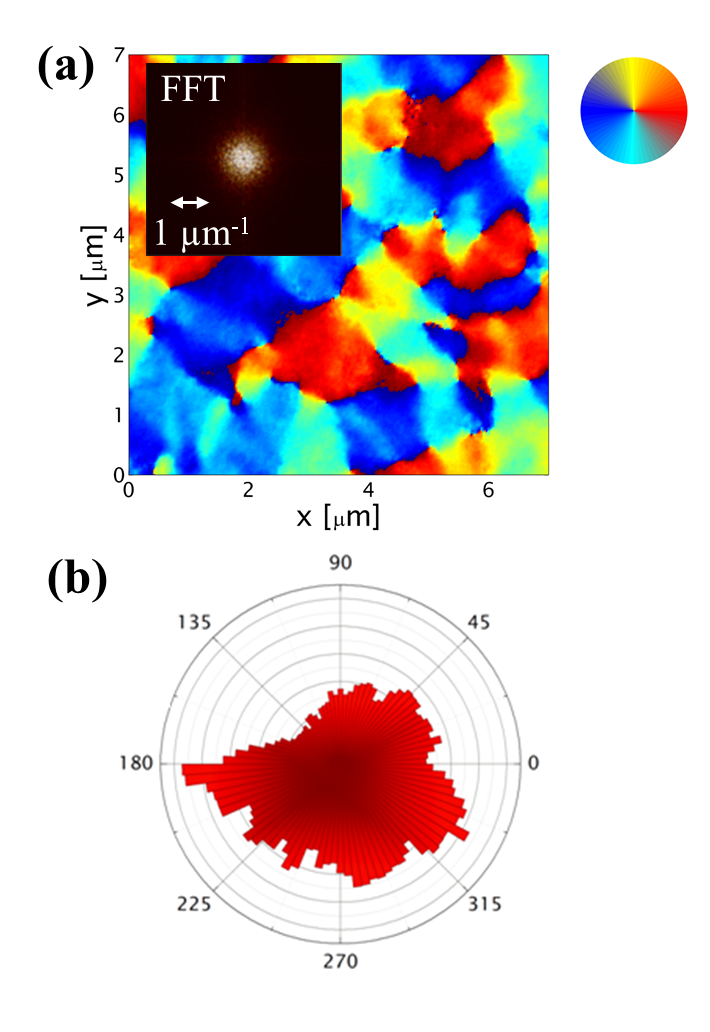}\\
\caption{(a) The vector image of the FM domain structure at a sample temperature of 342 K while warming, under zero applied magnetic field. The colour wheel indicates the moment direction. The inset shows the FFT of the image used to estimate the magnetic domain size. The domain size was found to be approx 1.2 ~$\mu$m  (b) The extracted angular dependence of the magnetization.}\label{Fig3}
\end{figure}

By acquiring two images rotated by $90^{\circ}$ with respect to each other it is possible to produce a vector map of the FM structure. To maximise the XMCD contrast in the images, the cubic [001] crystal axis of the FeRh(Pd) phase was aligned in the direction of the photon propagation vector. Figure~\ref{Fig3}(a) shows the vector domain distribution at 342~K upon warming, well inside the FM phase with zero applied magnetic field. As can be clearly seen the sample consists of micron-sized FM domains. Due to the angular dependence of the XMCD, out-of-plane magnetization results in a reduced XMCD contrast which can be seen in Figure~\ref{Fig3}, localised near the magnetic domain boundaries. In-plane magnetization is anticipated in the studied film. AFM domains do not generate any XMCD. From the image we can extract the angular dependence of the magnetization as shown in Figure~\ref{Fig3}(b). Recent measurements by Mariager \textit{et al.}\cite{Mariager_arXiv_2013} do not find evidence of a strong magnetocrystalline anisotropy in the undoped FeRh. We also observe only weak evidence of a magnetocrystalline anisotropy which would reflect the fourfold cubic symmetry of the crystal structure. The FFT of the vector map shown in the inset of Figure~\ref{Fig3}(a) yields a domain size of 1.2 ~$\mu$m in agreement with the MFM image in Figure~\ref{Fig2}(e).

As was noted in the MFM images, XPEEM domains do not appear to be coupled to structural inhomogeneity in the sample. It was also observed that the domain structure is stable on the time scale of hours at a fixed temperature.

\begin{figure*}
  \includegraphics[width=185mm]{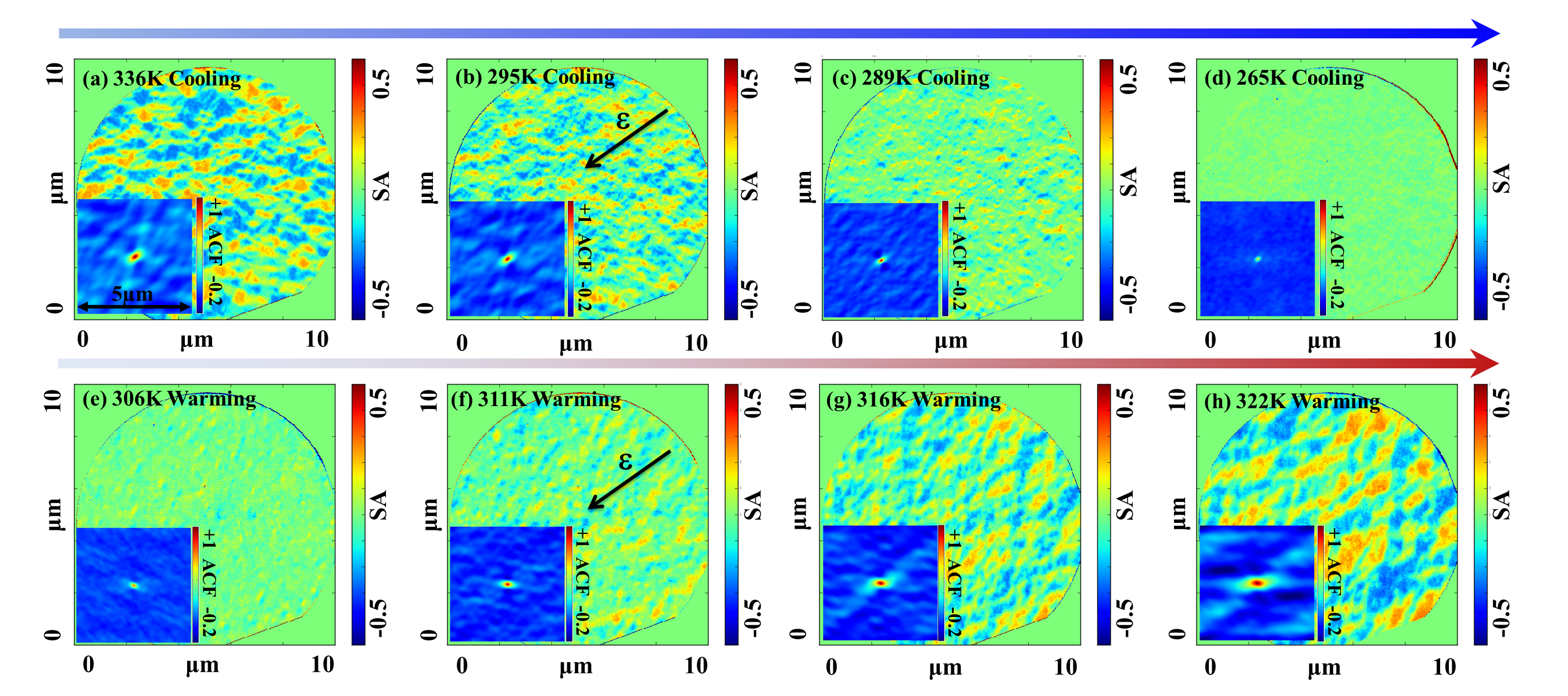}
  \caption{XPEEM images of the magnetic domain evolution of the Pd-doped FeRh film sample taken during progressive cooling, (a) to (d), and warming, (e) to (h), through the FM$\Longrightarrow $AFM and AFM$\Longrightarrow$FM transitions respectively. The XPEEM images were taken in zero applied magnetic field. The field of view is a 10~$\mu$m diameter. The arrows in panels (b) and (f) represents the photon propagation direction. The colour bar (SA - spin asymmetry) represents the normalized XMCD signal. The insets show 5~$\mu$m~$\times~5~\mu$m ACFs discussed in the text, with 1 and -1 being full correlation and full anti-correlation, respectively, and zero no correlation.}\label{Fig4}
\end{figure*}

Figure~\ref{Fig4} reveals the evolution of the magnetic domain configuration in the FeRh(Pd) sample with temperature as imaged by XPEEM. Starting in the FM phase (Figure~\ref{Fig4} (a)) regions of approximately micron-sized FM domains are visible, with polarization that is both parallel and anti-parallel to the cubic axis. Significant regions are aligned orthogonal to this direction (zero contrast); any AFM regions will also display zero contrast. As the sample is cooled (panels (b)$\rightarrow$(d)) the FM domains reduce in size with a concomitant increase in the regions of zero contrast. At 265~K (Figure~\ref{Fig4}(d)) there remains evidence of a weak FM component. The observation of low temperature FM regions in this temperature regime is qualitatively consistent with Ref.~\onlinecite{Fan2010}, which proposed a surface arrangement consisting of FM domains in an AFM matrix and with the XPEEM results of Baldasseroni \textit{et al.}\cite{Baldasseroni2012} for an Al-capped FeRh film sample. This result is also corroborated by the observation by Ding \textit{et al.}\cite{ding2008} where a small Fe L-edge XMCD signal at room temperature was observed originating in the near surface region of a similar FeRh film.

Warming back through the transition, it is noticeable that the magnetic domain sizes are significantly larger (by a factor of 2) than the size attained through cooling, as shown in Figure~\ref{Fig4} (e) to (h). Surprisingly, upon warming the domains are elongated along a cubic axis of the epilayer, which are seen at 45 degrees to the horizontal axis. This anisotropy in domain orientation is unusual and unexpected given the fourfold nature of the crystal symmetry.

In the transition region a reduced XMCD contrast is also visible between the domains. These regions are of a congruent shape but smaller size to the high contrast regions. In the case of a strong in-plane cubic anisotropy we would only expect three colour levels on the SA (spin asymmetry) images corresponding to magnetization $\mathbf{M}$ parallel (red) and antiparallel (blue) to $\bm{\varepsilon}$ and to orthogonal components or antiferromagnetic ordering (green). These regions between the (anti-)aligned domains presumably primarily contain antiferromagnetic ordering and domain walls. Finally, after warming above the coexistence regime we recover the fourfold-like structure (Figure~\ref{Fig3}(a)).

\begin{figure}
  \includegraphics[width=75mm]{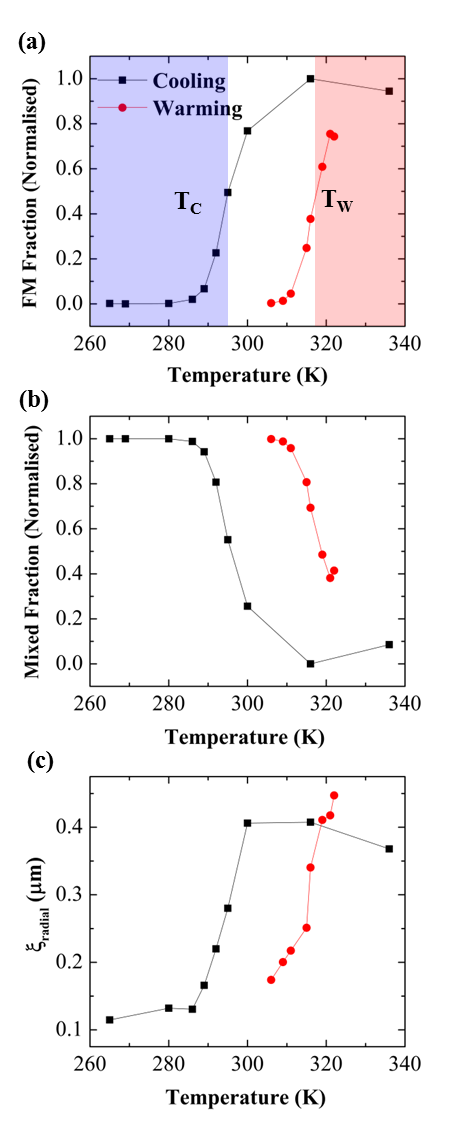}
  \caption{a) The hysteresis loops extracted from the XPEEM data. The normalized FM magnetization (summation of the SA) as a function of temperature is comparable to that observed by SQUID. The coloured panels represent the regions above and below the midpoint of the transition. b) The normalized AFM and orthogonal XMCD components (summation of the SA) as a function of temperature. Note the reversal of the hysteresis loop. c) The correlation lengths $(1/e)$ as a function of temperature extracted from the radial intensity plots.}\label{Fig6}
\end{figure}
\section{Analysis \& Discussion}

The XPEEM images allow extraction of the  temperature dependence of the magnetization by summing the XMCD contrast within an image, as shown in Figure~\ref{Fig6}(a). We label the mid-point of the AFM-FM transition for the cooling and warming cycles as $T_c$ and $T_w$ respectively. These XPEEM derived results are in good qualitative agreement with the sample-averaged magnetization derived from the SQUID measurements shown in Figure~\ref{Fig2}. There are two main differences, the first is a temperature offset of approximately 7~K between the XPEEM and the SQUID measurements, with the XPEEM data showing the onset of ferromagnetism before that of the SQUID. This might be ascribed to thermometry differences in the two techniques, however it has been shown that a reduced moment but persistent FM region near the cap/film interface approximately 60\AA\ thick exists as observed in Ref.~\onlinecite{Fan2010}. This region could induce the surface region into a FM state at a lower temperature than the bulk due to the effective exchange field. Secondly the width of the thermal hysteresis curve measured by the XPEEM technique, around 20~K, is approximately 10~K smaller than those measured by SQUID magnetometry.  As the XPEEM technique samples both the surface and near-surface ordering this region dominates the observed signal as compared to the sample averaged SQUID measurements.

A sum over the regions of zero XPEEM contrast provides an indication of the degree of AFM phase content as well as regions of orthogonal FM. This is shown in Figure~\ref{Fig6}(b) and we observe an inverted hysteresis loop as expected for the increase and decrease of the AFM phase with cooling and warming respectively.

To quantitatively analyze the phase character of the XPEEM images we have calculated the two-dimensional autocorrelation functions (ACFs)\cite{Sivia} of each real space image, displayed as insets in Figure~\ref{Fig4}. Autocorrelation is a signal analysis tool useful for extracting weak signals in rapidly varying noise, hence its use here in order to quantify domain size and orientation in the XPEEM images. Starting at high temperature (inset Figure~\ref{Fig4} (a)) and cooling, a cross-shaped fourfold symmetric structure is visible with well defined maxima corresponding to a FM domain size of $\approx 1.2 \mu$m in both the vertical $(y)$ and horizontal $(x)$ directions, indicative of a magnetic domain pattern with that underlying symmetry. As the system is cooled the FM domain signature disappears, and only the central peak is visible (\textit{i.e.} short range correlations as shown in the inset Figure~\ref{Fig4}(d)) as expected given the loss of XMCD contrast due to the appearance of the AFM phase.

The change in the ACF signal, and hence the magnetic domain behavior on warming the Pd doped FeRh film is more noticeable (inset Figure~\ref{Fig4} (e)). Starting at low temperature the central ACF peak is somewhat elongated. Upon warming into the FM phase the ACF central peak and satellites are anisotropic reflecting the elongation of the FM domain structure, resulting in an Ising-like nematic ordering with a reduced twofold symmetry. This effect is also visible in the real space image of the 322K Pd doped FeRh film (\textit{c.f.} Figure~\ref{Fig4} (h)).

As a consistency check, the normalized integrated intensities around concentric circles were calculated as a function of distance from the central point in the image, referred to as a radial intensity plot. Extracting the $1/e$ correlation length of this radial intensity, shown in Figure~\ref{Fig6}(c), recovers the same temperature dependence as that displayed by the magnetization of the Pd doped FeRh film. This method removes any angular dependence on the choice of cut direction, which is necessary since analysis of the diffuse background in 2D ACF images requires further development\cite{Sivia}.

\begin{figure}
  \includegraphics[width=92mm]{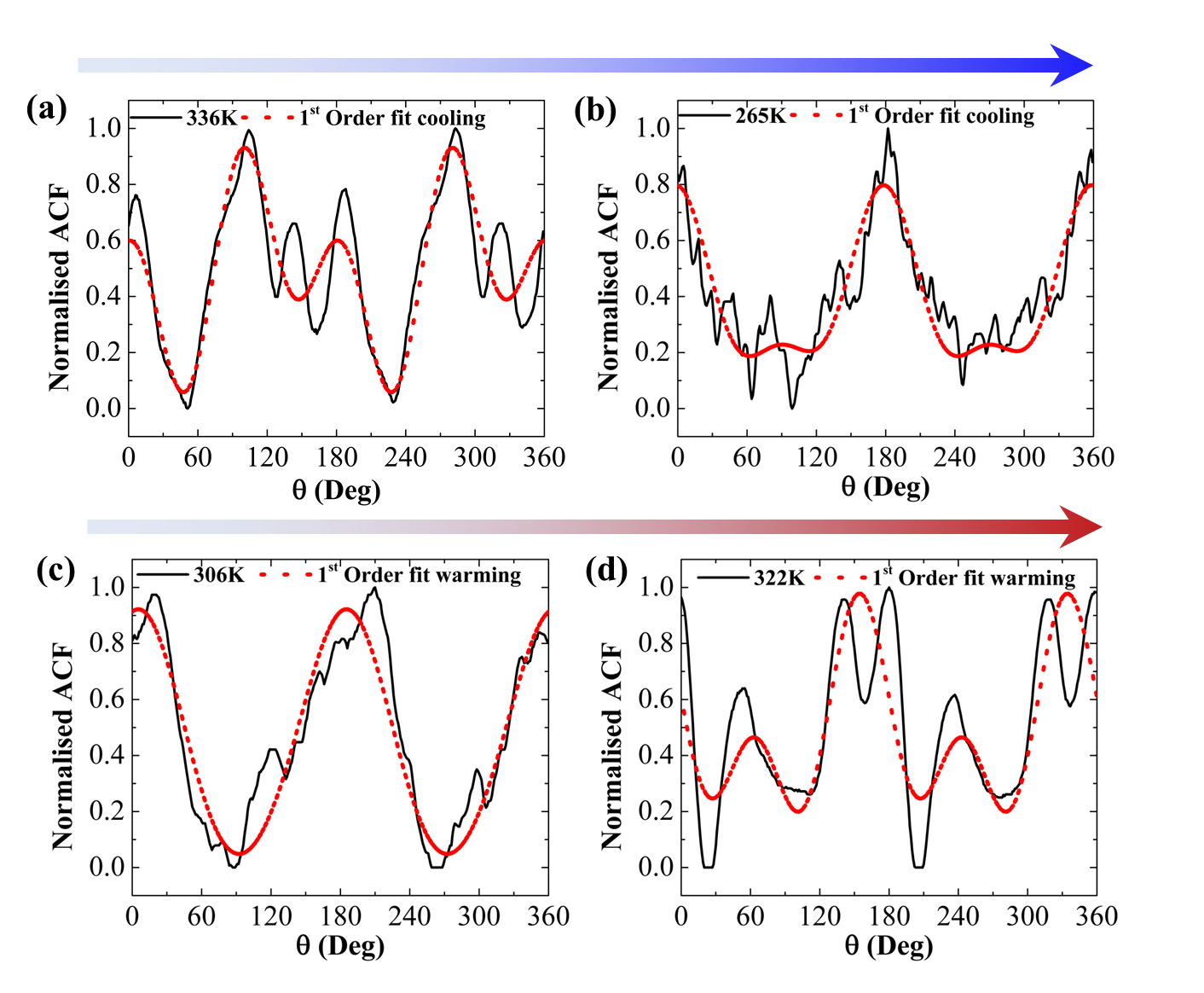}
  \caption{(a) The normalized and smoothed ACF as a function of the polar angle $\theta$ at a radius corresponding to the average domain size at 336~K on the cooling branch of the Pd-doped FeRh film. The $C_4$ symmetry is clearly visible. The dashed curve is the best fit of Eq. \ref{equ1}. (b) For the lower temperature of 265~K, the ACF is dominated by a twofold $C_2$ symmetry. Some remanence of the fourfold symmetry is also visible. Panels c) and d) show the warming case where the phase with a significant $C_2$ component moves towards a mixed $C_2$/$C_4$ state.}\label{Fig7}
\end{figure}

To quantify the change in symmetry of the magnetic character of the FeRh(Pd) film in warming and cooling through the first-order phase transition, a circular section through the four ACF satellite peaks as a function of polar angle $\theta$ for the 336~K and 265~K cooling images are plotted in Figure~\ref{Fig7} (a) and (b). The warming curves are shown on panels (c) and (d). The radius of these circular sections was determined from the ACF for each temperature. The central peak of the ACF is defined as $n=0$, where $n$ is the order of the peak as a function of increasing radius. The radius of the circular cuts used to generate the angular dependence was chosen to correspond to the nearest neighbor maxima in the ACF, $n=1$. In order to capture the magnitude of the different twofold $C_2$ and fourfold $C_4$ symmetry terms, we empirically described the circular sections with the following expression:
\begin{equation}
I(\theta) = C_{2} \cos^{2}(\theta + \phi_{2}) + C_{4} \cos^{2} (2\theta + \phi_{4}).
\label{equ1}
\end{equation}
where phase offsets between the twofold and fourfold symmetry terms are given by $\phi_2$ and $\phi_4$, respectively. Higher order terms in the series were neglected in the fitting. The fits are shown as dotted lines in Figure \ref{Fig7}.

\begin{figure}
  \includegraphics[width=75mm]{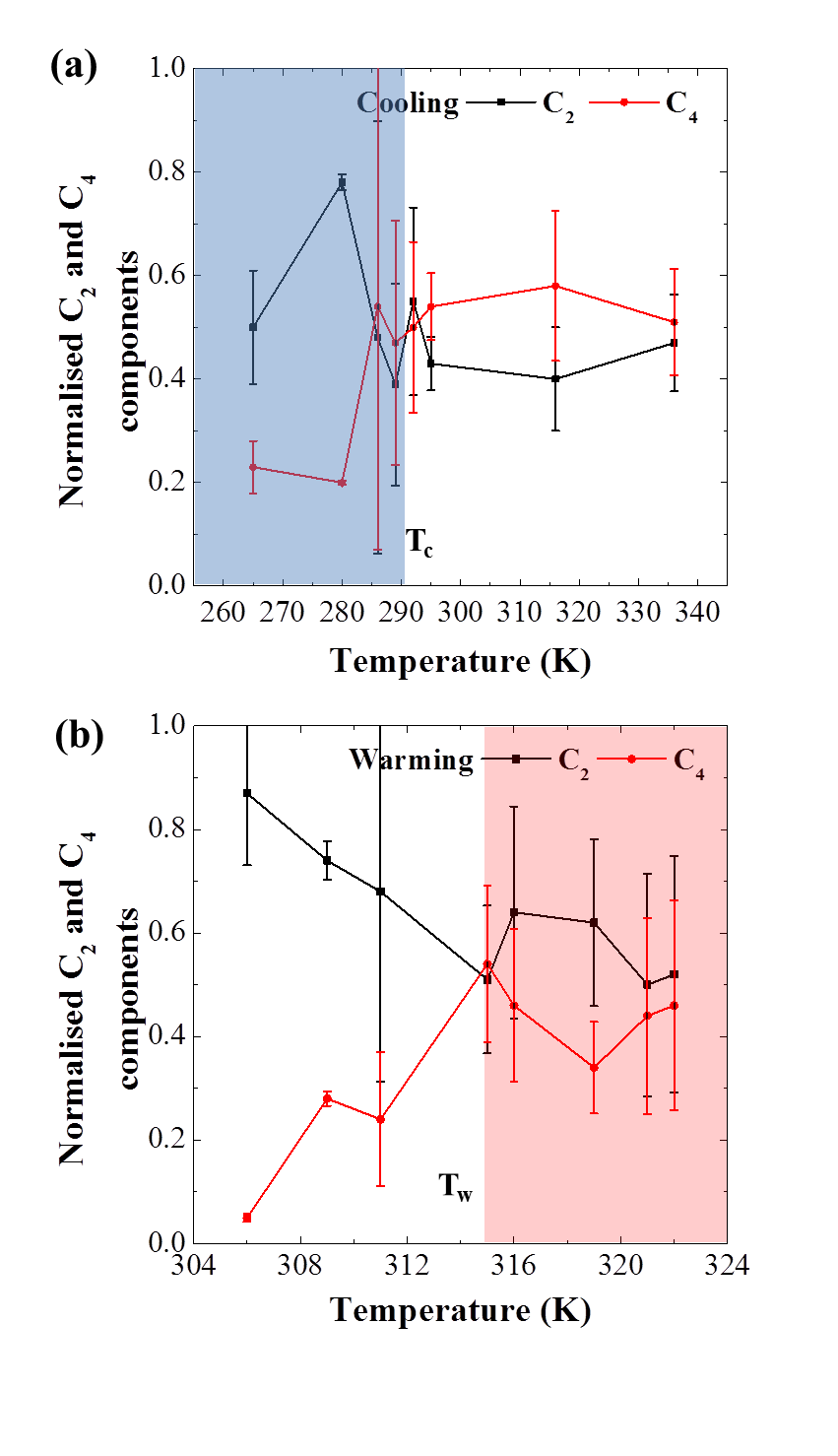}
  \caption{Temperature dependence of the $C_4$ and $C_2$ components obtained from fits to the circular sections of the ACFs for the (a) cooling and (b) warming cycles.}\label{Fig8}
\end{figure}

Figure~\ref{Fig8} shows the trend for the  $C_2$ and $C_4$ components for the temperatures at which XPEEM imaging was performed for both cooling, Figure 8(a) and warming, figure 8(b). In the high temperature region of the cooling panel there is an approximately equal balance of $C_4$ and $C_2$ symmetries. Below $\approx 290\textrm{K}$ on cooling and $\approx 315\textrm{K}$ on warming the relative balance changes to become dominated by the $C_2$ symmetry. From both the SQUID and XPEEM hysteresis loops these temperatures correspond to the mid-point of the transition from AFM$\Rightarrow$FM $(T_w)$ and FM$\Rightarrow$AFM $(T_c)$. No discernable trend with temperature was observed in $\phi_2$ or $\phi_4$. Variations in $\phi$ away from zero were allowed in the fitting to account for small changes in relative alignment during the XPEEM imaging.

The origin of the weak magnetic fourfold $C_4$ symmetry is consistent with the in-plane fourfold crystal symmetry (see Figure~\ref{Fig1a}(b)). Conversely, the mechanism leading to the lowered-symmetry $C_2$ term is more obscure. In doped FeRh the disorder created by the introduction of Pd into the lattice gives rise to broken translational symmetry, thereby locally modifying the degeneracy of the AFM state and the competing $C_4$ and $C_2$ FM ground states. Not only does quenched disorder exist in FeRh(Pd) but a significant elastic strain is present that is generated between the AFM and FM domains arising from the disparity in the AFM/FM unit cell volumes. Clearly, a large elastic strain exists in the undoped material but this does not appear to generate a change in magnetic domain symmetry. Indeed a recent systematic capping layer study\cite{Baldasseroni2014} appears to downplay the importance of strain in driving the transition. This suggests that the disorder is a significant driver for the observed change in magnetic symmetry. For higher Pd doping $(x\approx 0.1)$ it is known that the FeRh(Pd) system adopts a martensite, body centered tetragonal L1$_0$ type order. Such systems exhibit complex shape memory behavior\cite{Fukuda2013} whilst maintaining the AFM-FM transition. We have no evidence of such higher doping levels in our sample (XRD, EDX). However, it is conceivable that with the introduction of disorder and a surface (providing additional freedom for the relief of strain) that a similar change in symmetry could be possible at lower doping concentrations. Ref.~\onlinecite{Fukuda2013} shows that for $(x\approx 0.1)$ Pd doped FeRh has two different hysteresis curve widths depending on details of the sample's thermal history. A $C_2$ symmetry surface relief (stripes) phase was also observed in this work upon cooling, and is characteristic of a martensitic phase. Small regions of the stripe phase persisted up to high temperature. This would seem a reasonable driving mechanism for the $C_2$ symmetry component we observe in the XPEEM and its behavior with temperature.  

Further insight can be gained by making an analogy to the complex oxides. Calculations by Ahn~\textit{et al.}\cite{Ahn2004} on manganite perovskites have shown that long and short range anisotropic elastic distortions can give rise to self-organized stripy, $C_2$-type symmetry on the micron-scale and a change in the density of states resulting in either metallic or insulating behavior. Both localised (chemical doping) and long length scale (expanded/unexpanded lattice regions associated with FM/AFM domains) strain fields are present in Pd-doped FeRh.


The proximity of FM and AFM regions in the Pd doped FeRh films also suggests that an exchange bias may be present which would also serve to lower the rotational symmetry and the domain propagation process\cite{Fitzsimmons2000}. Interestingly, Su \textit{et al.}\cite{Su_PRL_2011} have recently reported on the emergence of rotational symmetry from disordered exchange biased systems reinforcing the role of exchange bias in the breaking of the rotational symmetry in the present system. In FeRh(Pd) little evidence of exchange bias is seen in the magnetometry. It is possible that any net exchange bias signal from the FM/AFM domain states would be very small and difficult to detect with conventional magnetometry. Hence the role of exchange bias in FeRh and FeRh doped films would seem to be a further avenue of inquiry using other techniques that are sensitive to microscale magnetism.

A final observation to be made is that for low temperatures (lower than the mid-point of the two transition temperatures $T\le T_c, T_w$) we recall that a remanent surface FM phase exists in capped epitaxial films. This, coupled with the near-surface sensitivity of the XPEEM technique suggests that the near-surface FM component has a reduced symmetry. The SQUID data in figure \ref{Fig2} when compared to the equivalent XPEEM data in figure \ref{Fig6} implies that the majority of the film is behaving differently. Hence we postulate that the relative influence of the $C_2$ phase diminishes as the temperature is increased and the bulk of the film transforms into the FM phase. This results in the domain structure reflecting the (weak) bulk-like $C_4$ symmetry.

\section{Summary}

To summarize, in a (3 at.\%)Pd doped FeRh thin film we have observed the coexistence of both antiferromagnetic and ferromagnetic order while warming and cooling through the magnetic transition. Quantitative analysis of the in-plane domain structure suggests a temperature-dependent change in symmetry: the expected $C_4$ symmetry lowers to $C_2$ in the phase coexistence region and dominates to lower temperature. This behavior contrasts to that found in the un-doped material. The origins of the different symmetries appears to be linked to the competing disorder present in the system resulting in a martensitic like near-surface phase coupled to the more bulk-like ordering within the majority of the epilayer. It is conjectured that this is analogous to the strongly correlated oxide systems, where even small amounts of doping can dramatically change the nature of the phase coexistence. The interaction of structural, electronic degrees of freedom and disorder results in a system that can be tuned to operate at room temperature with a stable, and controllable electronic/magnetic domains structure. The complexity of FeRh combined with chemical doping lead to interesting functional behavior with technological applications.

\begin{acknowledgments}
This work was supported by the EPSRC grant reference EP/G065640/1, Department of Energy Office of Basic Energy Sciences and the National Science Foundation (NSF) DMR-0908767 and DMR-0907007. We would like to thank Diamond Light Source Ltd for the provision of X-ray beamtime.
\end{acknowledgments}

\bibliography{ferh}

\end{document}